# Soliton groups as the reason for extreme statistics of unidirectional sea waves

A.V. Slunyaev[1)] and A.V. Kokorina

*Institute of Applied Physics, 46 Ulyanova Street, Nizhny Novgorod, Box 120, 603950, Russia, Slunyaev@hydro.appl.sci-nnov.ru, and Nizhny Novgorod State Technical University n.a. R.E. Alekseev. 24 Minina Street, Nizhny Novgorod, 603950, Russia*
[1)] ORCID iD 0000-0001-7782-2991

**Abstract**

The results of the probabilistic analysis of the direct numerical simulations of irregular unidirectional deep-water waves are discussed. It is shown that an occurrence of large-amplitude soliton-like groups represents an extraordinary case, which is able to increase noticeably the probability of high waves even in moderately rough sea conditions. The ensemble of wave realizations should be large enough to take these rare events into account. Hence we provide a striking example when long-living coherent structures make the water wave statistics extreme.

## 1. Introduction

Today it is generally recognized that unidirectional nonlinear sea waves may posses extreme statistical properties compared to the directional sea (in particular, Onorato et al, 2002, 2009; Gramstad & Trulsen, 2007; Waseda et al, 2009; Mori et al, 2011). As claimed by most of the studies, the reason for the crucial difference is the modulational (Benjamin – Feir) instability of deep water waves, which is much more efficient in the unidirectional regime. Though different group forms may exhibit unstable nonlinear dynamics, the criterion for instability of unidirectional uniform waves (the side-band instability) is the most recognized. The weakly nonlinear theory states that for waves with amplitude $A$ and wavenumber $k_0$ a wavenumber domain of long unstable perturbations $K$ exists,

$$0 < K \leq K_{BF}, \quad K_{BF} = 2\sqrt{2}\varepsilon k_0, \quad \varepsilon = k_0 A. \qquad (1)$$

The water wave steepness $\varepsilon$ cannot exceed some certain value due to the breaking mechanism (roughly speaking, $\varepsilon < 0.4$). Sea wind waves are typically characterized by the values $\varepsilon \sim 0.1$ or may be less, then unstable perturbations should be long (proportional to $\varepsilon^{-1}$, see (1)). The growth of the wave modulation occurs due to the coherence between consecutive waves in the train. The nonlinearity forces near-resonant wave quartets to interact efficiently. In this regime the Fourier harmonics exchange energy so that the modulation develops. A significant wave enhancement occurs when the train of coherent waves is long. In reality the incoherent wave component may distort or even prevent the focusing of long trains. Meanwhile, the modulational growth of the weakly nonlinear waves requires long time (or distance) to develop (proportional to $\varepsilon^{-2}$, since 4-wave interactions are responsible for this process), thus the modulated wave trains have to maintain the coherence in large domains in time and space, and only then they are able to grow significantly.

In the fully nonlinear numerical simulations of perturbed unidirectional uniform waves by Slunyaev & Shrira (2013) the initial condition with the steepness $\varepsilon \approx 0.1$ and larger break. This result approximately agrees with the in-situ observations that in sea states characterized by the steepness exceeding ~0.1 waves overturn from time to time (e.g. Babanin, 2011). The breaking phenomenon limits the maximum attainable wave amplification. Thus from the viewpoint of abnormally high waves caused by the modulational instability, moderately steep sea states are likely the most promising condition; this expectation is confirmed by some observations (Kharif et al, 2009). As mentioned before, the growth time for these waves is inversely proportional to the squared steepness; the fastest growth observed in the laboratory and numerical simulations of non-breaking waves takes a few tens of wave periods (e.g., Shemer & Sergeeva, 2009; Slunyaev & Shrira, 2013).

The nonlinear stage of the modulational instability may be approximated by breather solutions of the nonlinear Schrödinger equation (NLS). This framework allows a detailed investigation of the unstable modulation dynamics involving sophisticated mathematical tools (such as the Inverse Scattering Technique, IST, see Drazin & Johnson, 1996; Osborne, 2010), relatively simple analytic solutions are also available. There is a number of laboratory and numerical studies which show reasonable agreement between the predictions of the simplified NLS theory and the actual water wave dynamics (Chabchoub et al, 2011, 2012a,b; Shemer & Alperovich, 2013; Slunyaev et al, 2013b; Slunyaev & Shrira, 2013). The breather solutions of the NLS equation are strongly localized disturbances of the wave envelope, which start to grow from very small perturbations and eventually may exceed many times the amplitude of the background wave. They satisfy the conventional definition of a rogue wave, that it should exceed the characteristic wave height at least twice. More exactly, in oceanography the rogue wave criterion usually reads (Kharif et al, 2009)

$$\frac{H}{H_s} > 2, \qquad (2)$$

where $H$ is the wave height according to the zero-crossing analysis of the time series, and $H_s$ is the significant wave height, which is frequently defined as $H_s = 4\eta_{rms}$, where $\eta_{rms}$ is the root-mean-square of the surface displacement $\eta$, or as the averaged over one third of the highest waves, $H_s = H_{1/3}$ (see e.g. Massel, 1996).

Within the NLS framework a breather may be treated as a superposition of an envelope soliton and a background plane wave (see details in Slunyaev, 2006, and also Akhmediev & Ankiewicz, 1997). The envelope soliton is another exact solution of the NLS equation, which is a stationary wave group which does not disperse with time and interacts elastically with other waves. In terms of the IST, one soliton and one breather are both characterized by a single discrete eigenvalue, but the background conditions are different: zero for an envelope soliton and a plane wave for a breather (Soto-Crespo et al, 2016). When the amplitude of the background wave decreases, the breather solution tends to the envelope soliton. The recent laboratory investigations and numerical strongly nonlinear simulations confirm that soliton-like groups in deep water may indeed propagate stably (Dyachenko & Zakharov, 2008; Slunyaev, 2009; Slunyaev et al, 2013a, 2017), and this situation is not confined by the weakly nonlinear range, where the NLS theory is formally valid. Steep solitary wave groups are short, having a few oscillations in a snapshot. The maximum steepness of the solitary wave groups registered in laboratory conditions reach the value $\varepsilon \approx$ 0.3 Slunyaev et al, 2013a, 2017); then the wave crests are close to the limiting Stokes wave shape. Such solitary groups obtained in the numerical simulations are shown in Fig. 1 at six different wave phases (the white curves).

Breathers represent 'elements' of the dangerous unstable wave dynamics; a NLS breather is a superposition of an envelope soliton and the background (Akhmediev & Ankiewicz, 1997). The idea to detect breathers or solitary groups in the time series of water waves before they cause danger (and hence to make the warning of extreme waves possible) is rather attractive and is being implemented in different ways (Osborne, 2010; Islas & Schober, 2005; Slunyaev, 2006). The unstable modes or solitary wave groups were revealed with the help of the IST in instrumental records of the extreme events in the in-situ and laboratory conditions. Strongly localized wave groups and individual waves which live for tens of wave periods are reported in publications in relation to the rogue wave phenomenon, see for examples the accidents with the naval ship Jeanne d'Arc with the Glorious Three, and with research vessel Ballena in California (Kharif et al, 2009).

Due to the growing computer performance and the development of numerical algorithms for the integration of hydrodynamic equations, numerical simulations are considered as a promising way to obtain the reliable probabilistic description of the sea waves in fully controlled conditions (e.g., Janssen, 2003; Socquet-Juglard et al, 2005; Onorato et al, 2009; Shemer et al, 2010; Chalikov, 2009; Zakharov & Shamin, 2010; Slunyaev & Sergeeva, 2012; Sergeeva & Slunyaev, 2013; Xiao et al, 2013; Slunyaev et al, 2016; and others). Unidirectional deep-water waves with large Benjamin – Feir Index (introduced in Onorato et al, 2001; Janssen, 2003)

$$BFI = \sqrt{2} \frac{k_p \eta_{rms}}{\Delta \omega / \omega_p} \qquad (3)$$

are prone to generate extreme waves. In the limit of weak nonlinearity and slow modulations the Benjamin – Feir instability develops if $BFI > 1$. Then the wave height probability distribution function (PDF) significantly exceeds the reference Rayleigh distribution (valid for linear narrow-banded waves, see e.g. Massel, 1996) when waves are high. In (3) $k_p$ and $\omega_p$ are the peak wavenumber and frequency linked by the deep water dispersion relation $\omega_p^2 = gk_p$, $g$ is the gravity acceleration, and $\Delta\omega$ is the width of the frequency spectrum.

The most important theoretical achievements from the standpoint of the relation between the spectral and statistical properties of irregular waves were obtained by Mori & Janssen (2006) and Onorato et al (2016) for stationary and transient situations respectively, governed by the weakly nonlinear theory for narrow-banded waves. They show that the fourth statistical moment for the surface displacements, the kurtosis, may increase anomalously if $BFI$ is not small, and that the variation of the spectral shape (which is observed when $BFI > 1$) unavoidably results in the change of kurtosis. In particular, for the NLS equation for unidirectional waves in deep water broadening of the spectrum leads to the increase of kurtosis.

The instability criterion (3) for irregular waves is related with condition (1) for a modulated plane wave. Irregular waves are unstable ($BFI > 1$) when for a given characteristic wave steepness $k_p\eta_{rms}$ (which cannot exceed some realistic value) the spectrum is sufficiently narrow. The issue of how to evaluate the characteristic spectrum width for realistic sea waves with heavy-tail spectra and spectra of complicated shapes is not obvious. When the famous JONSWAP spectrum is considered,

$$S(\omega) = \alpha \left(\frac{\omega}{\omega_p}\right)^{-5} \exp\left[-\frac{5}{4}\left(\frac{\omega}{\omega_p}\right)^{-4}\right] \gamma^{\exp\left[-\frac{1}{2\Delta^2}\left(\frac{\omega-\omega_p}{\omega_p}\right)^2\right]}, \qquad (4)$$

$$\Delta(\omega) = \begin{cases} 0.07, & \omega < \omega_p \\ 0.09 & \omega > \omega_p \end{cases},$$

where $\alpha$ is constant, large values of the peakedness $\gamma$ or very large wave heights are required to meet the instability condition. In particular, in strongly nonlinear simulations of the potential Euler equations by Sergeeva & Slunyaev (2013) two sea states characterized by different values of $H_s$ exhibited remarkably different statistical properties. The one with $H_s \approx$ 3.5 m, $\gamma = 3$ and peak period $T_p \approx 10$ s (sea state A) possessed the PDF rather close to the Rayleigh law, though the other with $H_s \approx 7$ m and $\gamma = 3.3$ (sea state E) exhibited much larger probability of high waves. During the simulation of series E numerous occasional wave breaking occurred, what required regularization of the related numerical instability.

The increased probability of high waves in the situation E was associated with long-living coherent wave groups hidden in the irregular waves. The groups could be observed in the time-space plots of the surface displacement (in fact, in the both series A and E); the evaluated lifetimes of rogue events were up to a few tens of the wave periods. Besides, some difference in the time dependencies of the instantaneous *BFI* was pointed out in Slunyaev (2010) for the situations of initially large, moderate and small values of *BFI*. The curves *BFI*(*t*) calculated over moderately large ensemble of realizations in situations *BFI*(0) > 1 experienced weak oscillations at time $t/T_{nl} > 1...2$, $T_{nl} = \omega_p^{-1}(k_p \eta_{rms})^{-2}$, with characteristic time of the oscillations exceeding the wave period. This observation could be explained by the emergence of the coherent group structures. However, we could only speculate on whether the localized structures indeed exist, and if so, whether they correspond to breathers or solitary groups.

In this paper we represent an example when a soliton-like wave group emerges in an irregular sea state of unidirectional waves characterized by the JONSWAP spectrum with a moderate peakedness $\gamma = 3$ and a moderate wave height $H_s \approx 3.5$ m, $T_p \approx 10$ s. An ensemble of 300 wave realizations calculated in the space-time domain of the size 10 km × 20 min is considered. The intense short wave group occurs without any clear precursor and dramatically changes the wave height probability distribution function. We discuss possible physical effects leading to this dynamics.

## 2. The initial condition and the approach for data collecting and processing

The initial conditions for the direct numerical simulations are prepared in the form of surface displacements $\eta(x, t = -200 \text{ s})$ and surface velocity potentials $\Phi(x, t = -200 \text{ s})$ related according to the linear solution for waves in infinitely deep water (the first 200 seconds of the simulations are reserved for the preliminary adjustment stage, see below). The Fourier transforms of the surface displacement are prescribed by the JONSWAP function (4) with $T_p = 10$ s, $4\eta_{rms} \approx 3.49$ m, $H_{1/3} \approx 3.37$ m and $\gamma = 3$, and random phases. Each realization is represented by a wave train of the length 10 km, what for the given wave parameters corresponds to about 60 dominant wavelengths. The waves evolution is calculated with the help of the High-Order Spectral Method applied to the potential Euler equations (West et al, 1987), what leads to the nonlinear boundary conditions at the water surface, $z = \eta(x, t)$,

$$\frac{\partial \eta}{\partial t} = -\frac{\partial \Phi}{\partial x}\frac{\partial \eta}{\partial x} + \left(1+\left(\frac{\partial \eta}{\partial x}\right)^2\right)\frac{\partial \varphi}{\partial z}, \qquad (5)$$

$$\frac{\partial \Phi}{\partial t} = -g\eta - \frac{1}{2}\left(\frac{\partial \Phi}{\partial x}\right)^2 + \frac{1}{2}\left(\frac{\partial \varphi}{\partial z}\right)^2\left[1+\left(\frac{\partial \eta}{\partial x}\right)^2\right]. \qquad (6)$$

In (5), (6) the velocity potential $\varphi(x, z, t)$ satisfies the Laplace equation, $\nabla^2 \varphi = 0$, and decays when $z \to -\infty$. The problem is periodic in space. The vertical gradient of the potential $\partial \varphi / \partial z$ is calculated on the surface; this quantity is calculated approximately with the use of the high-order asymptotic procedure which ensures taking into account up to 7-wave interactions (the HOSM nonlinearity parameter $M = 6$). This method corresponds to the almost fully nonlinear approach.

The resolution in space is about 130 grid points per dominant wave length; the stepping in time is 200 iterations of the 4-th order Runge–Kutta scheme per one wave period. The aliasing is removed for each pairwise product by virtue of calculation in the Fourier domain of a doubled size.

The preliminary stage $-200 \leq t < 0$ is used for the purpose of a smooth adjustment to the nonlinearity, the nonlinear terms of the equations are put in force slowly; this approach prevents excitation of the spurious small-scale harmonics, see Dommermuth (2000). The wave fields tend to achieve a quasi-stationary state (Sergeeva & Slunyaev, 2013). The results of the wave evolution within the next 20 minutes, $0 \leq t \leq 1200$, are recorded with the time interval about 0.6 s. The 20-min duration corresponds to the time which takes the wave with a period 10 s to pass the distance of 10 km. Meanwhile it corresponds to the typical length of the time series used in the oceanography, in our case it is about 120 wave periods. The typical maximal relative deviation of the total energy in the simulations is limited by the value $O(10^{-5})$. Thus, we deal with an accurate solution of the primitive potential hydrodynamic equations.

The ensembles of two sizes are considered in this paper: a small of $N_R = 100$ realizations and a larger of $N_R = 300$ realizations. Every simulation results in 2048 time series retrieved at different locations of the computational domain. The individual waves are selected in the time series according to the up-crossing approach; the down-crossing analysis leads to the similar results. The overall accumulated from different locations statistical data consist of up to $N = 614400$ samples ($N = N_R \times 2048$) what results in almost $10^8$ individual waves; note that the time series at close locations are treated as independent, though they are not. Thus the obtained statistics is not precisely the conventional one-point statistics; it rather corresponds to the statistics under an interval oriented along the direction of the wave propagation, and estimates the space-time probability. At the same time the correlation distance is likely less than the computational domain size of 10 km. The dedicated analysis of the correlation between waves registered at close locations performed in Sergeeva & Slunyaev (2013) showed that the correlation decreases as the wave nonlinearity grows. Therefore the consideration of the time series retrieved at different locations in one realization as independent seems to be a coarse but reasonable approach for improving the probabilistic description.

In this paper we use the Fourier transform in time or space for the estimation of the frequency or wavenumber spectra correspondingly. The frequency Fourier transform for the time interval $0 < t < 1200$ s averaged over 2048 locations and $N_R = 100$ simulations is shown in Fig. 2 in linear and semi-logarithmic scales (red solid lines), and compared with the JONSWAP curve used for the generation of the initial condition (broken lines). It is clear that the simulated waves and the JONSWAP spectrum well agree.

Assuming that the ensemble of the wave heights consists of $N_H$ values which are regarded as a random variable, let us rank them in decreasing order $H_1 > H_2 > ... > H_{NH}$. Then the estimator for the wave height exceedance probability distribution function is

$$P(H_m) = \frac{m}{N_H + 1}, \qquad m = 1, 2, ..., N_H, \qquad (7)$$

see Fig. 3 for the ensemble of 100 simulations. This curve is rather close to the Rayleigh distribution defined by

$$P_R(H) = \exp\left[-2\left(\frac{H}{H_s}\right)^2\right]. \qquad (8)$$

In Fig. 3 the simulated heights are scaled versus the quantity denoted by $H_s$, which is put equal to either $4\eta_{rms}$ or $H_{1/3}$. It is clear from the figure that the results of the simulations are even closer to the theoretical curve (8) in the small amplitude range if the quantity $H_{1/3}$ is used for the reference. The proximity to the Rayleigh law is observed despite the facts that: i) not all the time series used for the statistical processing are statistically independent; ii) the effect of wave nonlinearity is important; iii) the spectrum is wide; and finally iv) the statistical ensemble is not infinite. A small upward excursion from the Rayleigh curve is observed below probabilities about $10^{-4}$, when $H_s = H_{1/3}$. In Sergeeva & Slunyaev (2013) a smaller ensemble was considered ($N_R = 20$); no noticeable extreme statistical properties of the waves with $H_s \approx 3.5$ m were found either (see below Fig. 4a).

## 3. The extraordinary wave realization

The probability distribution function for a larger ensemble of 300 realizations was found to be noticeably different in the tail from the PDF for $N_R = 100$ (Fig. 4a): the probability of high waves is significantly larger than the reference Rayleigh curve. Note that the deviation between the curves for $N_R = 20$, $N_R = 100$ and $N_R = 300$ starts from about the same level of probability, $\sim 1 \cdot 10^{-3}$. The stability of estimation of the exeedance probability for the data of finite size we evaluate with the help of the standard deviation $\sigma$ for random heights $H_m$ (see e.g. Tayfun & Fedele, 2007)

$$\sigma = \frac{1}{N_H + 1}\sqrt{m\frac{N_H - m + 1}{N_H + 2}}, \qquad m = 1, 2, ..., N_H. \qquad (9)$$

The areas $P \pm \sigma$ are shown in Fig. 4 with shading. The comparison between the PDFs calculated for the first, second and third hundred of realizations from the set of 300 simulations (Fig. 4b) shows noticeable difference between the curves for probabilities less than $1 \cdot 10^{-3}$.

After a thorough investigation the 'extraordinary' realization No 295 was revealed with the maximal wave; it is characterized by the PDF which exceeds the Rayleigh distribution at the probability level above $10^{-2}$ (dash-dotted curve in Fig. 4a). The initial condition for this extraordinary realization is shown in Fig. 5a. The maximum displacement in Fig. 5a corresponds to $\eta_{max}$ = 3.14 m, or $\eta_{max}/4/\eta_{rms}$ = 0.90; the ensemble averaged value $\eta_{rms}$ = 0.87 m is used for the analysis. The maximum crest-to-trough vertical distance is $H_{max}/4/\eta_{rms}$ = 1.65. Seemingly, it has no specific peculiarities which could definitively result in the subsequent extreme dynamics and statistics; except, may be, the presence of a long wave train of moderate amplitude in the middle of the domain. The frequency spectrum for this realization (averaged over 2048 spatial locations) is shown in Fig. 5b. The spectrum generally follows the wanted JONSWAP shape, though some irregularity due to the nonlinear exchange between the spectral components is obvious.

The results of a more involved investigation of this simulation are discussed below. The top-view on the surface displacement in the space-time domain is presented in Fig. 6b. There the system of references is used, which is co-moving with the group velocity of the dominant waves, $C_{gr} = \omega_p/k_p/2$. The maximum surface displacement during the standard period of the numerical simulation, $0 \leq t \leq 120T_p$, is attained at time $t_* = 1042$ s $= 104T_p$ and coordinate $x_* = 4766$ m $= 31\lambda_p$, where $\lambda_p = 2\pi/k_p$ is the wave length; the maximum displacement is characterized by the value $\eta_{max}$ = 6.58 m, or $\eta_{max}/4/\eta_{rms}$ = 1.88. The evolution of the global maximum $\max_x|\eta(x, t)|$ as a function of time is given in Fig. 6c, where the filled circle denotes the moment $t = t_*$. The surface in Fig. 6b is centered at the coordinate $x = x_*$. It is clear that the intense waves in the area $|x - x_* - C_{gr}(t - t_*)| \lesssim 3\lambda_p$ contain major part of the wave energy in the entire domain of simulations. As already mentioned, the group in the centre of Fig. 6b is characterized by moderately large waves from the very beginning, $t = 0$, having the maximum wave amplitude about $4\eta_{rms}$; it further increases with time up to about $7.5\eta_{rms}$ at $t = t_*$ (Fig. 6c). Traces of other transient wave groups are readily seen in Fig. 6b; some of them propagate faster than the dominant waves and move rightward in the used references; slower trains drift leftwards. The smaller-amplitude groups live shorter than the 'central' intense wave group.

According to Fig. 6b several stages of the large wave group evolution may be suggested.

1) A slow convergence of waves from $t = 0$ to $t \approx 60T_p$.

2) A strong transformation of the group preceding the maximum wave formation, $60T_p < t < 100T_p$. The stage, when a broad wave group is shortening and increasing the amplitude looks very like the modulational instability; the rough estimation of the parameters of the group at $t = 0$ with respect to the criterion (1) reveals that the group might be modulationally unstable.

3) A quasi-stationary propagation of a solitary group ($t > t_*$). At about $t \approx t_*$ a very localized steep wave group appears, which preserves the shape till $t = 120T_p$ almost unchanged. A longer numerical simulation of the realization was performed beyond $120T_p$, until $t = 245T_p$, with the purpose to observe the further evolution of the intense group. The longer simulation reveals that the group remains to a large degree stable and retains the energy during the next ~100 periods (i.e., $120T_p < t < 220T_p$). Meanwhile a significant variability is observed seemingly due to the interaction with other waves (in particular, see $t \approx 80T_p$ and $t \approx 150T_p$). Shifts of the maximum group location are observed at about these instants (Fig. 6b). The maximal surface displacements seem to occur shortly after the collisions between the waves (Fig. 6c).

The evolution of the spatial Fourier transform of the waves shown in Fig. 6 is given in Fig. 7. It is clear that initially the spectrum is unimodal, it inherits the feature of the initial condition. With time, one smooth spectral hump breaks into many local peaks, though two close intense peaks seem to accumulate most of the energy, they are well-seen from $40T_0$ to $120T_0$. (We would like to mention that the wavenumber spectrum obtained through the averaging of 300 realizations does not manifest any clear evolution.)

The window spectra of a few snapshots of the wave evolution are presented in Fig. 8 (the squared Fourier amplitudes are plotted). The corresponding instants are marked with broken white lines in Fig. 6. The window spectra in Fig. 8 display the spatial distribution of the energy held by different wave scales. The white curves in Fig. 8 represent the corresponding surface displacements. The sampling window length is about $L_w = 20\lambda_p$, the smoothing Hanning mask is applied (see in Massel, 1996)

$$M(x) = 0.5\left(1 - \cos 2\pi \frac{x}{L_w}\right); \qquad (10)$$

the mask shapes are given in red in the upper left corners of the plots in Fig. 8. The spectrum evolution for the sampling window centered at $(x - x_* - C_{gr}(t - t_*)) = 0$ is displayed in Fig. 6a; this figure allows the selective consideration of the evolution of the most intense wave train.

Similar window spectra are given in Fig. 1 for different phases of the intense envelope soliton on a zero background. Despite the complicated strongly nonlinear dynamics of the steep solitary groups (the white curves in Fig. 1), in the space-wavenumber domain they are well defined by single spots of energy regardless the wave phase.

Initially the local wavenumber spectrum of the most intense wave area is one-humped (Fig. 6a), though the local frequency varies with coordinate (Fig. 8a). A second peak of the wavenumber spectrum starts appearing later on slightly ahead the most intense group (Fig. 6a, Fig. 8b). The two peaks seem to concur at about the moment of the maximum wave formation (Fig. 8c). At later stages, several wavenumber spectrum peaks are present at about the same spatial location (Fig. 8d-f). According to Fig. 6a, the peak wavenumber varies with time. It shifts to a smaller wavelength at $t \approx 150T_p$ and returns to longer waves at $t \approx 200T_p$. At the late stage a new – less intense, but steep wave group which propagates slower than the dominant waves may be noticed, see Fig. 6b for $t > 200T_p$ and the corresponding surface profile in Fig. 8f. Note the similarity between the short intense groups in Fig. 8c-f and in Fig. 1.

## 4. Discussion

In this paper we consider an example of strongly nonlinear simulations of irregular waves governed by the potential Euler equations. The wave realizations are prepared in accordance with the classic Longuet-Higgins (1952) assumption that sea waves are in the leading order sinusoids with random phases; they are characterized by the realistic JONSWAP spectrum with typical parameters of steepness and peakedness which correspond to moderately rough sea conditions not favourable for the modulational instability. The waves obey the assumption of planar geometry; more particular, the waves may be considered to a large degree unidirectional (the space-time spectra which reflect the waves propagating in different directions were discussed in Slunyaev & Sergeeva, 2012).

The unidirectional waves are known to exhibit extreme statistical properties, and in this work we explore if they can be attributed to the long-living nonlinear wave groups of the soliton kind. As the waves in the simulations are moderately steep and correspond to relatively wide spectrum, the observation of pronounced soliton-like structures was not anticipated originally. The wave height probability distributions calculated for moderate statistical ensembles (Sergeeva & Slunyaev, 2013; Slunyaev et al, 2016), see Fig. 3, 4a, did not deviate significantly from the Rayleigh function.

In a larger ensemble of the simulations an example is found, when an intense soliton-like wave group emerges; it contains most of the energy in the calculation domain, and preserves the energy noticeably longer than other wave groups. As a result, the realization No 295 exhibits extraordinary statistical properties, as the wave dispersion which shuffles the wave phases is suppressed.

With the purpose to enlarge the volume of statistical data the time series retrieved from different but close locations are considered as independent, though their correlation is evidently not negligible. When the intense solitary group emerges and propagates for abnormally long distance, the time series obtained at different locations become strongly correlated even if they are distant, what may interfere significantly the probabilistic properties of the wave heights (see review in Beran, 1992). Let us estimate the variability of the PDF associated with different choice of the location, where the time series are retrieved. When we imply that the time series from different locations are statistically independent, they form the common statistical ensemble with the probability $P$ plotted in Fig. 3,4. These exceedance probability distributions are reproduced in Fig. 9a,b by different lines for $N_R =$ 100 realizations (31 million waves), $N_R =$ 300 realizations (93 million waves), and for the realization No 295 (316 thousand waves, see the legend). Similar to Fig. 4, the bands of the standard deviation $\pm\sigma$ are also shown in Fig. 9 by the dark shading.

Without relying on the assumption about the statistical independence of the data retrieved from different locations, only one time series from each simulation may be included in the statistical database. Then, depending on the location $x_j$ where the time series are collected (assuming the location $x_j$ is the same for all the realizations), $N_{SS} =$ 2048 subsets of the data is available, each consists of $N_R$ time series. Each of the subset represents an alternative ensemble of statistically independent data. Consequently, $N_{SS} =$ 2048 alternative wave height probability distribution functions may be calculated, $P_j(H)$, $j =$ 1, …, $N_{SS}$. In every subset there are about from 15 000 to 46 000 individual waves for $N_R =$ 100 and $N_R =$ 300 correspondingly. The variability of the calculated exceedance probability with respect to the location $x_j$ may be assessed with the help of the standard deviation $\Delta$ for given $H$ defined as

$$\Delta(H) = \sqrt{\frac{1}{N_{SS}} \sum_{j=1}^{N_{SS}} (P(H) - P_j(H))^2} \ . \qquad (11)$$

In (11) $P_j(H)$ corresponds to the location $x_j$, and $P(H)$ is the exceedance probability (7) based on the data collected from all the grid points. The light-shadowed areas in Fig. 9a are confined by the curves $P + \Delta$ and $P - \Delta$ for $N_{SS} =$ 2048; they are much larger than the bands $P \pm \sigma$ shown by the darker colors.

Generalizing the two approaches, let us consider $Q$ time series retrieved in one simulation at the most distant locations of the computational domain as statistically independent. Then, depending on the locations, $N_{SS} =$ 2048/$Q$ alternative PDFs may be obtained. Each of the alternative ensembles will contain $N_R \times Q$ samples. The stability of the estimate for the exceedance probability with respect to variation of the location may be evaluated with the use of (11). The light-shadowed areas in Fig. 9b characterize the case $N_{SS} = 256$, i.e., when each $Q = 8$ time series from the most distant locations in one simulation are assumed independent. In other words, in Fig. 9b we assume that the data subsets collected at the distance about 1 km possess statistical independence. In Fig. 9a the light-shaded areas are nested according to the growth of the size of the statistical data. In Fig. 9b each of the subsets contains larger number of wave heights, and the uncertainty, caused by the choice of the locations where the data are retrieved, decreases. In Fig. 9b the probabilities estimated for the series 295 and for large ensembles look dissimilar.

The estimates for relative errors $\sigma(H)/P(H)$ and $\Delta(H)/P(H)$ are compared in Fig. 10 (curves with symbols versus plane curves) for the same numbers $N_R$ and $N_{SS}$ as in Fig. 9. The curves with circles denote $\sigma/P$ calculated for the data collected from all the grid points, and they are much below the other curves due to the large number of measurements. The subsets contain smaller number of the wave heights, and thus are characterized by different standard deviations $\sigma_j$. The subsets are non-overlapping, and thus the probability distributions $P_j$ may noticeably differ for large wave heights. The curves with triangles in Fig. 10 correspond to the functions $\sigma_j/P_j$ calculated for two representative subsets. The up-looking triangles denote the subset which contains the maximum wave among all samples. The down-looking triangles correspond to the opposite case, i.e., the maximum wave in this subset is the smallest among the highest waves in the subsets.

The rarest event in an ensemble (the highest wave) is characterized by the uncertainty as much as 100%, according to (7) and (9). For $N_{SS} = 2048$ (Fig. 9a) the variation of $\Delta/P$ at the level $H/H_{1/3} = 2$ is estimated as about 40% and 20% for the ensembles of 100 and 300 simulations respectively. When 8 time series from each simulation are used in one subset ($N_{SS} = 256$, Fig. 9b) these estimates drop down to about 10% and 5%. When the ensembles of 100 or 300 realizations are considered, the variations of probability between different subsets, $\Delta$, (plane curves) agree reasonably well with the standard deviations, $\sigma_j$, for the selected subsets (curves with triangles). However, in the case of the simulation No 295 the variation between subsets (i.e., dependence on the location) is smaller than the estimate for the error due to the finite sampling, especially in Fig. 10b, where time series from 8 most distant locations were regarded as independent. This observation may suggest that the distance of wave correlation in the series 295 is anomalously large.

The illustration in Figs. 9,10 suggests that the difference between the major part of the statistical data and the realization No 295 is indeed outstanding, much beyond the typical variability. It is significantly to mention that the curve $P(H)$ for $N_R = 300$ almost coincides with the PDF for even larger wave ensemble $N_R = 1000$ (not discussed in this study) at probabilities above $5 \cdot 10^{-4}$, and they rather well agree at lower probabilities.

It may be realized that the Longuet-Higgins (1952) concept of the sea waves, which was used to generate the initial conditions for the numerical simulations, may not always be suitable from the physical point of view. For example, the situation when all the waves in the space series are in phase is theoretically allowed and in principle may occur as one of the random realizations. For the conditions of the present simulations (i.e., for the given JONSWAP spectrum, the wave intensity and the size of the spatial domain) this would lead to the initial condition in the form of a localized within a few wavelengths group with the maximum crest height about 23 meters and with the negligible surface displacement in the rest 10 km (Fig. 11). This maximum displacement can be never surpassed for the given length of the spatial domain, $L$, and given wave spectrum. It is easy to understand that the maximum of surface elevation of the focused waves relates to the spatial domain as $\propto L^{1/2}$ (see the maximum waves for different domains in Fig. 11); thus the nonlinearity of the allowed co-phased waves increases when $L$ grows; at the same time the wave dispersion remains similar since the wave spectrum shape is constant.

The enormously high-amplitude wave train localized within a few periods in a 10 km basin can nether occur in reality as a result of a natural wind wave dynamics. If waves tend to focus in such group, most likely they would break long before they reach the maximum amplitude. No breaking events were detected among the first 100 simulations performed for the study, though in 4 cases among the next 200 realizations numerical instability resulted in the calculation interruption due to the occurrence of too steep waves. Thus, we may anticipate that random initial conditions *in a sufficiently large domain* will result from time to time to a wave breaking *for any value* of the significant wave height, what is a side-effect of the Longuet-Higgins concept. Generalizing this way of thinking, one may conclude that a set of unphysical initial conditions generated with the help of the Longuet-Higgins representation may exist in a large basin. Meanwhile the initial condition for the simulation No 295 looks ordinary; its evolution does not cause wave breaking. Therefore it is difficult to attribute its dynamics to any unnatural artifact of the approach for the Monte-Carlo-type simulations.

No adequate methods exist for the analysis of the soliton-like coherent wave structures which are discussed in the present paper. The nonlinearity is essential, thus linear approaches are inaccurate or not valid at all. The considered case No 295 is definitively beyond the approximations of weak nonlinearity and weak dispersion. The dynamics is not described by the integrable nonlinear Schrödinger equation, the groups are not solitons in the strict mathematical sense; they can exhibit inelastic interactions and thus the Inverse Scattering Technique is also inapplicable. In this study we use the advanced window Fourier analysis to reveal the details of the wave group evolution. The wavelet analysis was previously applied to the in-situ records of rogue waves by, e.g., Chien et al (2002) and Mori et al (2002). In the present study the full space-time dynamics is available, and thus a more detailed investigation may be performed. The analysis is still unable to lead to doubtless conclusions; some statements may be made, though.

1) Firstly, the abnormally intense wave group in the simulation No 295 was not produced originally as a random superposition of sinusoids (which, as discussed above, may in principle create unrealistic wave forms). The waves could propagate for more than 240 wave periods without overturning.

2) The soliton-like short group of steep waves is seemingly created as a result of a self-modulation process of a long wave train. This process took about 60-80 wave periods. The emerged group stands out against the other waves; it accumulates most of the energy contained in the simulation domain. It leads to the extreme wave staistics.

3) After the intense group had been created, it propagated for at least ~120 wave periods as an individual solitary structure. Interactions with the other intense wave groups of similar wave lengths caused further wave height increase (including the generation of the maximum observed surface displacements), altering of the carrier wavenumber and fast shifts of the group localization. The feeding of one solitary wave by other waves (the 'champion soliton' effect, e.g. Pickartz et al, 2016) is not observed with evidence, but inelastic interactions in principle could help the large group to retain the energy.

4) The occurrence of an abnormally intense long-living wave group from the waves of moderate steepness exhibits the features of the 'black swan' effect. It could not be expected from the previous consideration of smaller statistical ensembles (Fig. 9), but when happens, produces an enormous effect on the wave statistics. Partly due to the method used for the statistical ensemble collection, this case not only increases the probability of very rare events, but noticeably influences the probability of moderately frequent events as well.

The persistence of soliton-like groups may play an important practical role in view of the rogue wave forecasting. Several approaches based on the proximity of the wave group dynamics to the NLS framework have been suggested recently (Slunyaev, 2006, 2017; Cousins & Sapsis, 2016), which theoretically should be able to track the modulationally unstable patterns in stochastic fields (e.g. Chabchoub, 2016). It is significant to emphasize in the end that the reported study concerns unidirectional waves. In the 3D case wave defocusing occurs in the transverse direction, which should weaken the self-modulation in the longitudinal direction, what may influence the dynamics essentially.

**Acknowledgement**

The research was supported by the Russian Foundation for Basic Research (grants No. 16-55-52019, 17-05-00067), the President grant for leading scientific schools SC-6637.2016.5, and also by the Volkswagen Foundation (for AS).

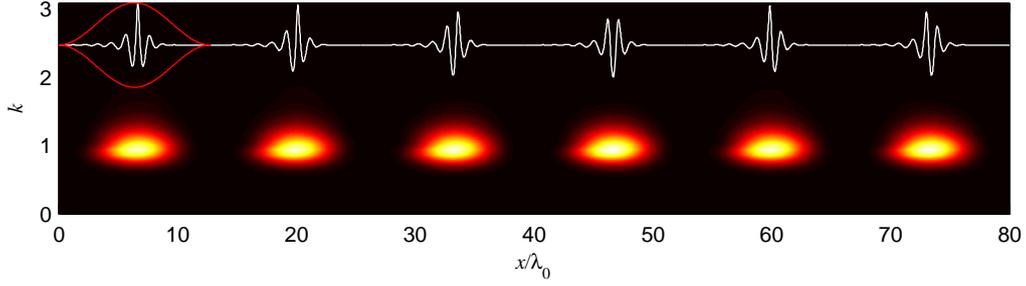

**Fig 1** Solitary wave groups with the steepness $\varepsilon \approx 0.3$ at six different phases in strongly nonlinear simulations of the potential Euler equations (white curves plot the surface displacement), and the window Fourier transform (the color intensity). Parameter $\lambda_0$ denotes the dominant wavelength. The sampling window with the Hanning mask (10) applied has the length $L_w$ about $13\lambda_0$ (see the mask shape given by red in the upper left corner)

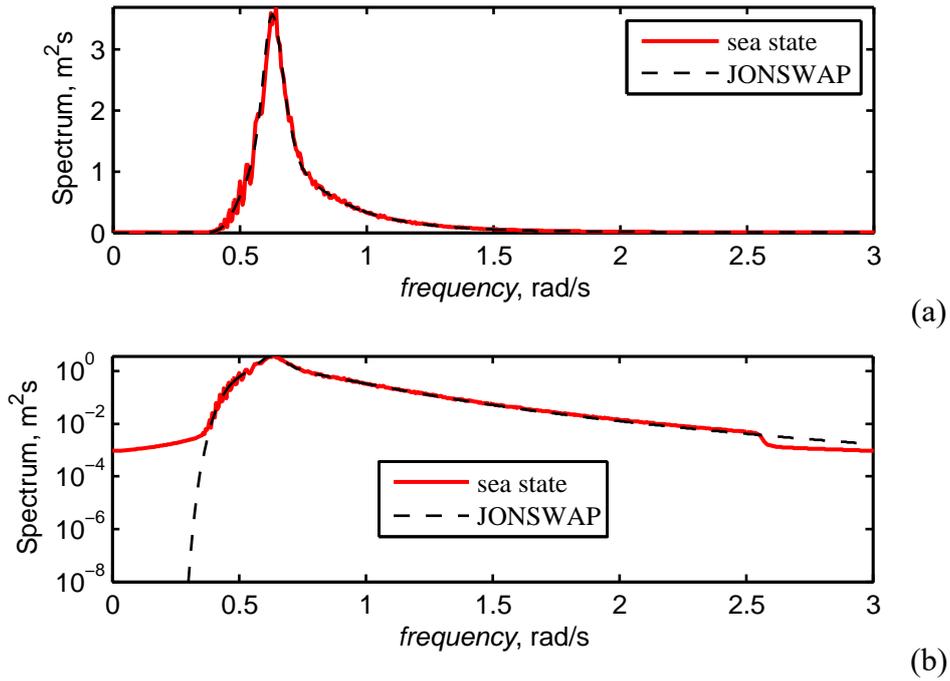

(a)

(b)

**Fig. 2** Averaged frequency spectrum compared with the wanted JONSWAP function (in the linear (a) and semilogarithmic (b) scales) for the ensemble of 100 realizations

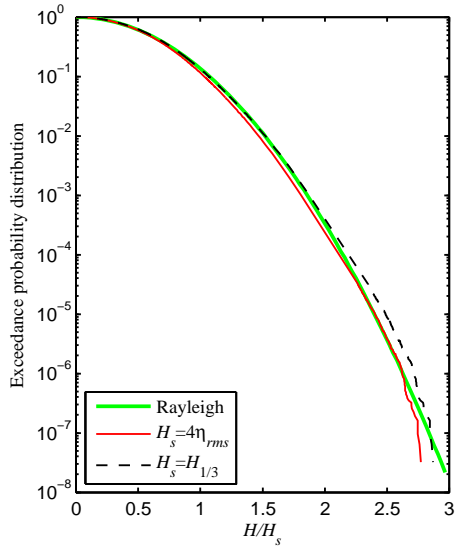

**Fig. 3** The wave height exceedance probability distributions for 100 realizations versus the wave height scaled by the values of $4\eta_{rms}$ and $H_{1/3}$. The reference Rayleigh distribution is also shown

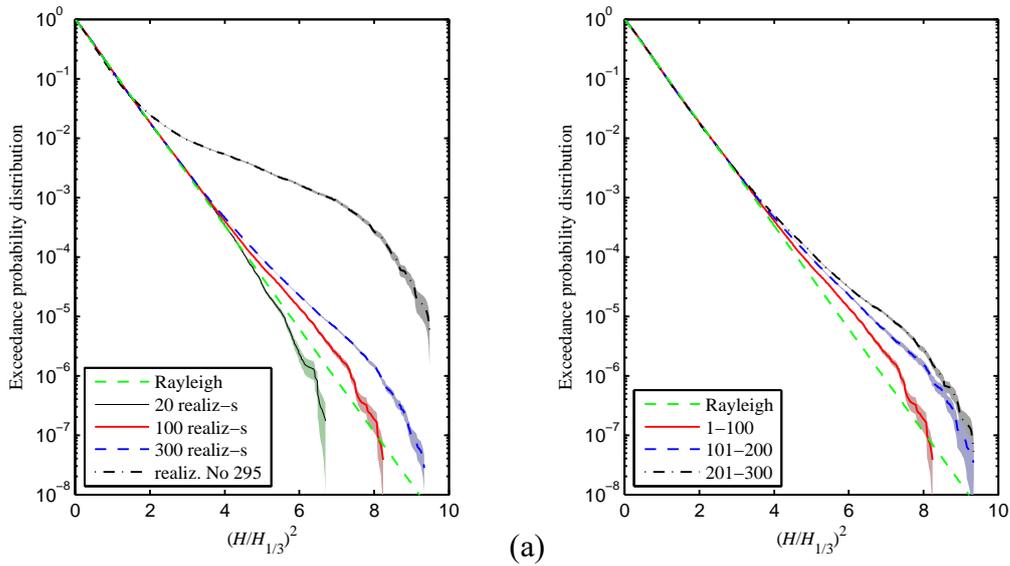

**Fig. 4** (a): The wave height exceedance probability distributions for the ensembles of $N_R$ = 20, 100 and 300 realizations, and for the extraordinary realization No 295 alone. (b): The PDFs for three sets of simulations from the database of 300 realizations. The standard deviation bands $\pm\sigma$ are given by shaded areas

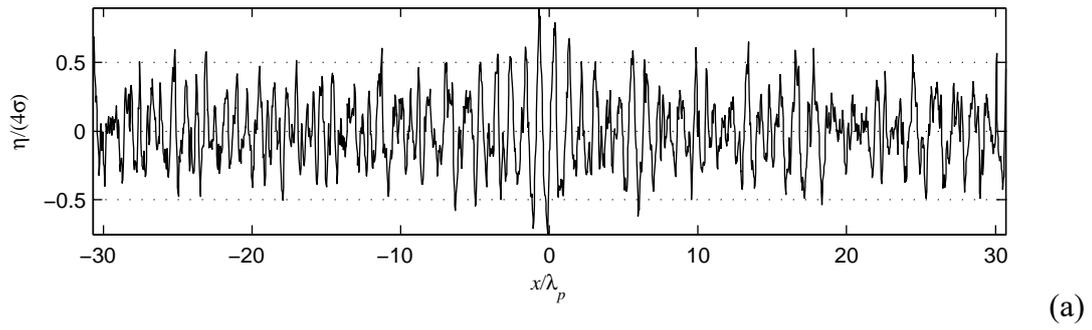

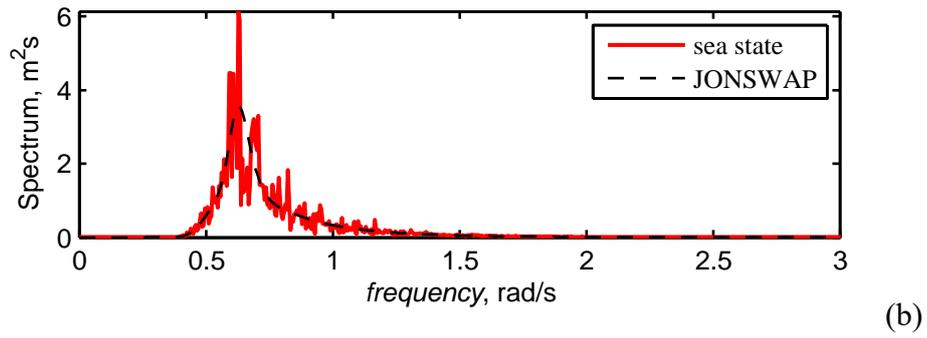

**Fig. 5** The extraordinary realization No 295: the initial condition (a) and the averaged frequency spectrum compared with the prescribed JONSWAP shape (b)

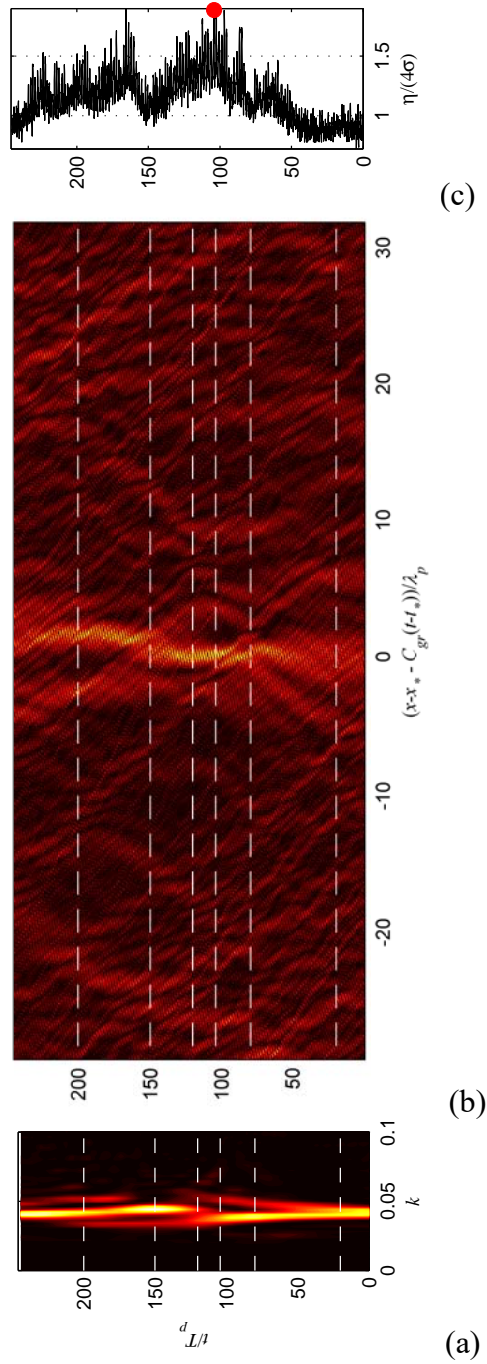

**Fig. 6** The extraordinary realization No 295: the top-view on the water surface in the commoving references (b), the local wavenumber spectrum for the Hanning data mask (10) with the width $L_w = 20\lambda_p$ centered at $(x - x_* - C_{gr}(t - t_*)) = 0$ (see the mask shapes in Fig. 8) (a), and the global maximum of the surface displacement $|\eta|$ as the function of time (c). The white broken lines in (a) and (b) correspond to the instants analyzed in Fig. 8; the red filled circle in (c) shows the moment of the maximum displacement attained in the interval $0 < t < 120T_p$

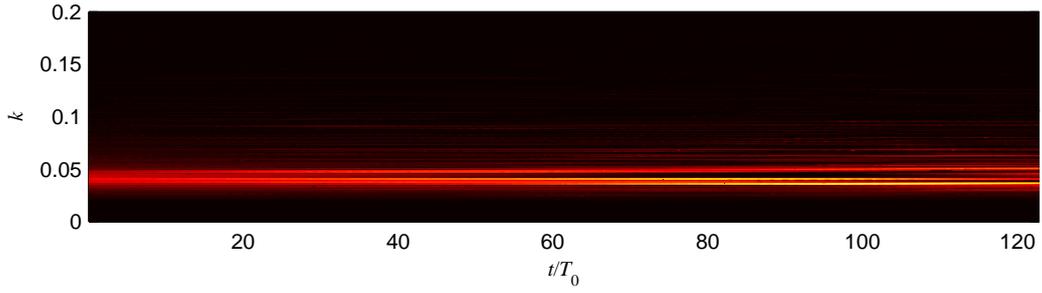

**Fig. 7** Evolution of the spatial Fourier transform (squared amplitudes) in time for the extraordinary realization No 295

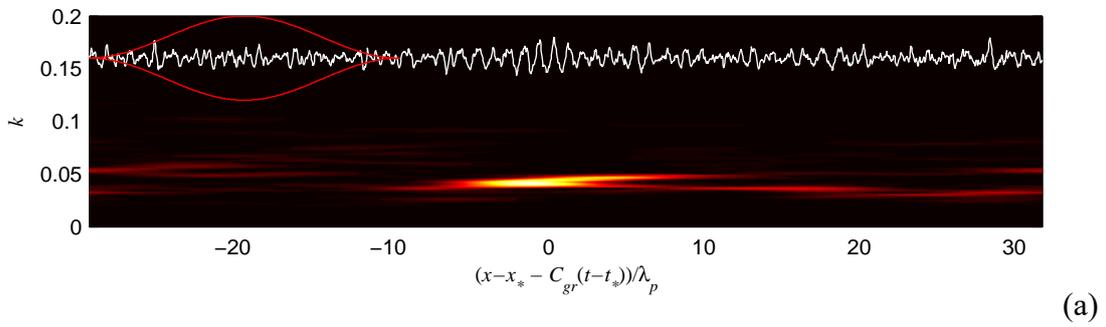

(a)

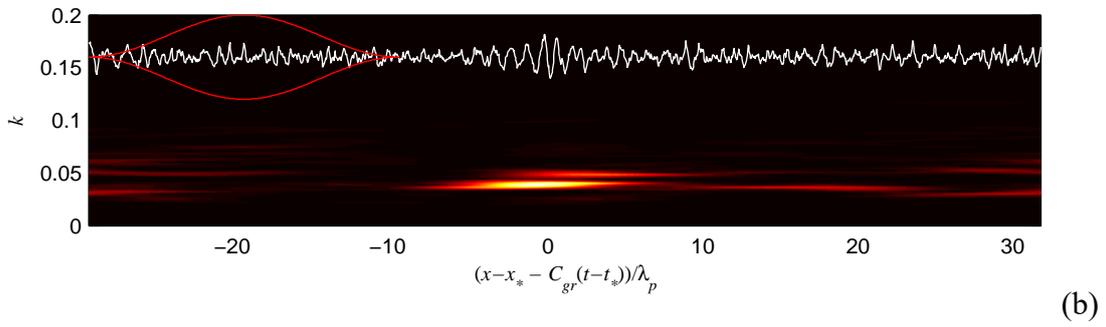

(b)

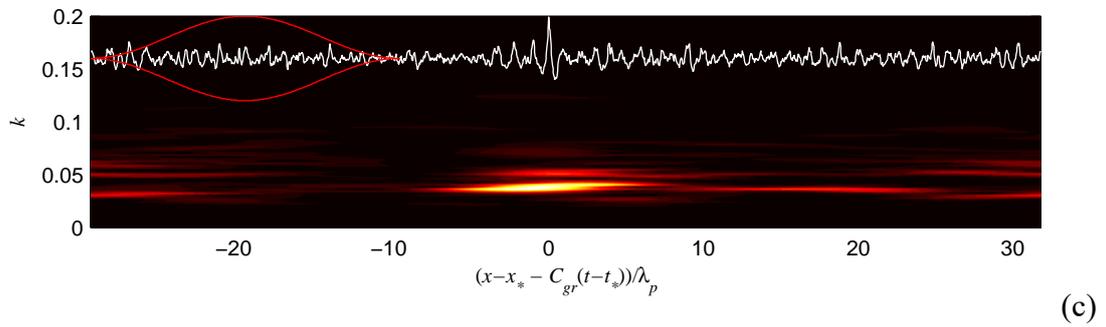

(c)

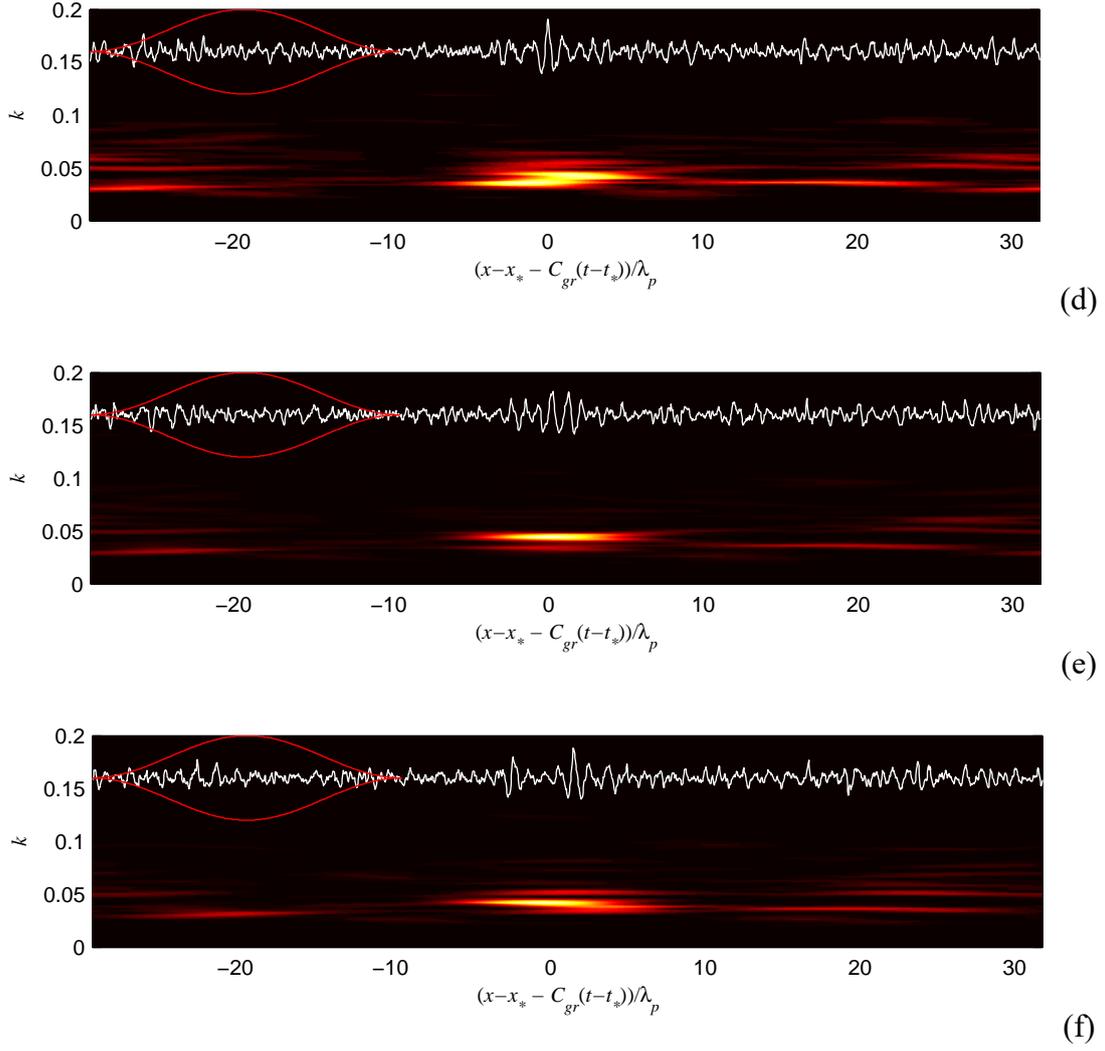

**Fig. 8** Window Fourier transforms for the extraordinary realization No 295 for instants $t = 20T_p$ (a), $t = 80T_p$ (b), $t = t_*$ (c), $t = 120T_p$ (d), $t = 150T_p$ (e), and $t = 200T_p$ (f). The sampling window length is about $20\lambda_p$ with the Hanning mask (10) (see the mask shape given by red curves in the upper left corners). The same intensity color map for the Fourier amplitudes is used in all the panels

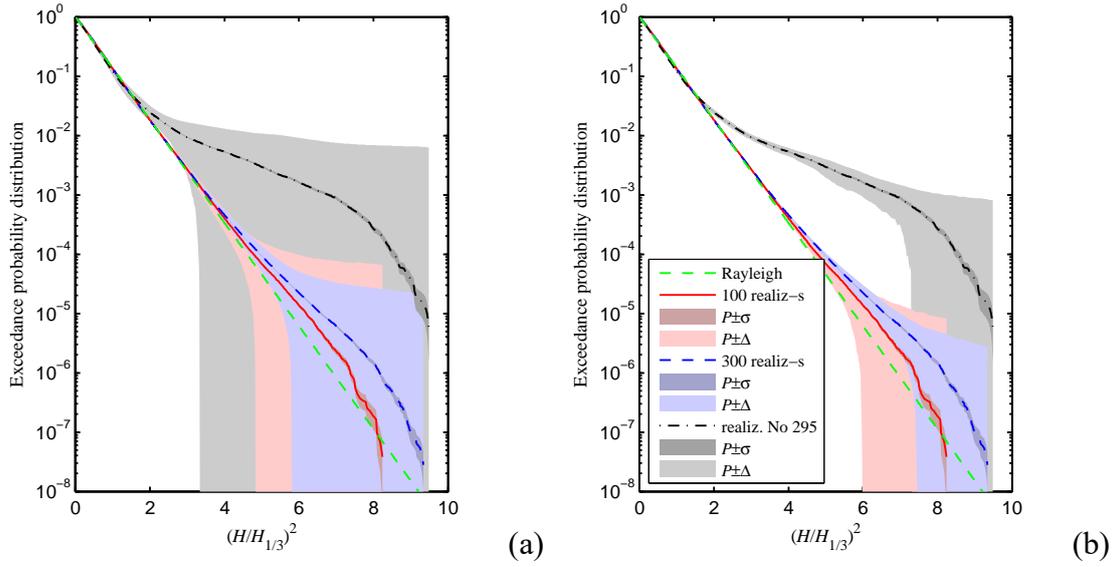

**Fig. 9** The wave height exceedance probability distributions for the ensembles of $N_R = 100$ and $N_R = 300$ realizations, and also for realization No 295. The shadowed areas correspond to the $\pm\sigma$ bands (dark colors) and the $\pm\Delta$ bands (light colors). The variation due to the different locations, $\Delta$, is estimated basing on $N_{SS} = 2048$ data subsets (i.e., when one time series from each realization is used) (a); and on $N_{SS} = 256$ subsets (i.e., when in each simulation 8 time series at 8 most distant locations are regarded as statistically independent) (b)

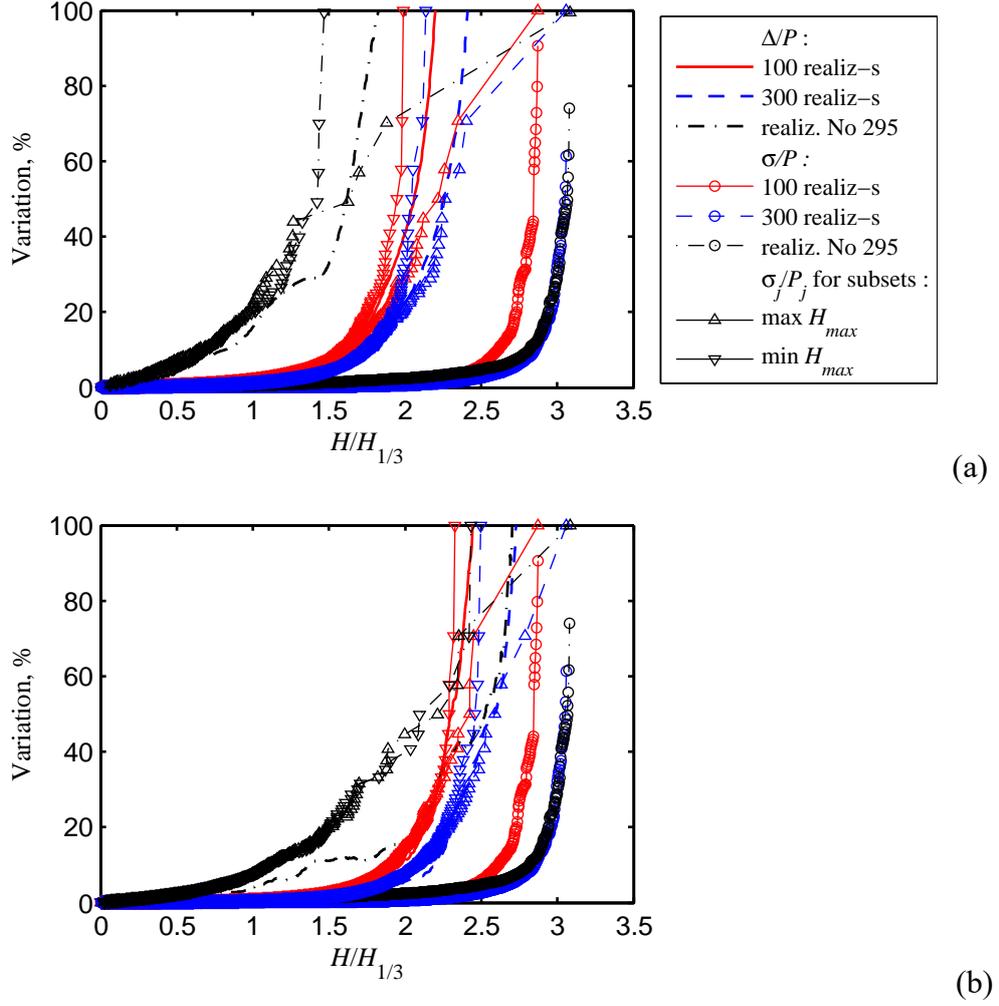

**Fig. 10** Estimates of the variability of the wave height exeedance probability for the same parameters $N_R$, $N_{SS}$ as in Fig. 9a,b. The plane curves represent $\Delta/H$, the curves with circles are $\sigma/P$ when all the time series are treated as independent. The curves with up- and down-looking triangles indicate relations $\sigma_j/P_j$ for the subsets with the maximal and minimal values of the largest wave heights, correspondently

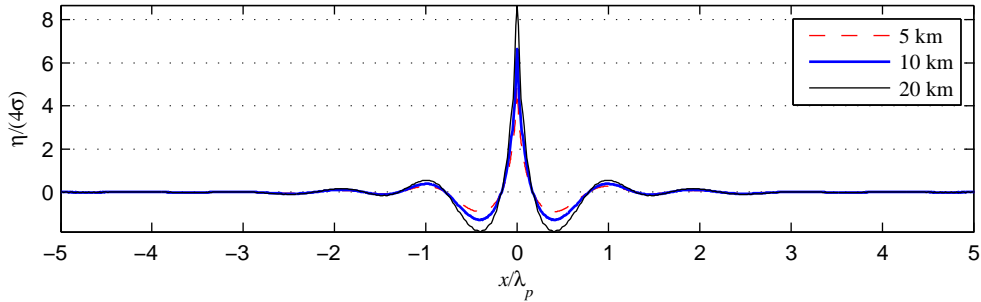

**Fig. 11** Fully in-phase waves with the given JONSWAP spectrum and $\eta_{rms} \approx 0.87$ for different sizes of the simulation domain $L$ (see the legend). Note that only a small part of the domain is shown